\documentclass[showkeys,aps,pre,reprint,superscriptaddress,showpacs,preprintnumbers,amsmath,amssymb,nofootinbib]{revtex4-2}
\usepackage{graphicx}% Include figure files
\usepackage{dcolumn}% Align table columns on decimal point
\usepackage{bm}% bold math
\usepackage{siunitx}
\usepackage{multirow}

\begin{document}

\preprint{}

\title{Design Optimization of Noise Filter using Quantum Annealer}

\author{Akihisa Okada}
\email{a-okada@mosk.tytlabs.co.jp}

\author{Hiroaki Yoshida}
\author{Kiyosumi Kidono}

\affiliation{%
TOYOTA CENTRAL R\&D LABS., INC., Bunkyo-ku, Tokyo 112--0004, Japan}%

\author{Tadayoshi Matsumori}
\author{Takanori Takeno}
\author{Tadashi Kadowaki}

\affiliation{%
DENSO CORPORATION, Minato-ku, Tokyo 108--0075, Japan}%

\date{\today}% It is always \today, today,

\begin{abstract}
The use of quantum annealers in black-box optimization to obtain the desired properties of a product with a small number of trials has attracted attention. 
However, the application of this technique to engineering design problems is still limited.
Here, we demonstrate the applicability of black-box optimization with a quantum annealer to the design of electric circuit systems, focusing on $\pi$-type noise filters as an example.
We develop a framework that uses quantum annealing to find the optimal location of electrical components and conductor paths connecting the components, and confirm that the learning process appropriately works over a number of trials to efficiently search for a design with high performance.
The results show the potential applicability of quantum annealing to design problems of electric circuit systems.
\end{abstract}

\maketitle

\section{Introduction}\label{sec:introduction}

\subsection{Quantum annealers}
High-performance computers are required to elucidate and predict complex phenomena, such as in simulations of the behavior of systems with multiple interconnected factors.
However, Neumann-type computers, whose development has followed Moore's law, do not meet the demand for high performance. Drastic improvements in Neumann-type computers are not expected~\cite{Moore2006} as their single-threaded performance has reached its ceiling~\cite{Moore2011}. Therefore, non-Neumann-type computers are expected to be an alternative for high-performance computation for complex problems.

Quantum annealers are one type of non-Neumann-type computer. Commercial machines are available from D-Wave Systems.
The architecture of a quantum annealer implements the Ising model on a circuit using superconductivity. The ground state of the Ising model is efficiently found using the quantum effect~\cite{Kadowaki}.
Since the ground state of the Ising model is equivalent to the solution of quadratic unconstrained binary optimization (QUBO), which includes not only fundamental problems~\cite{lucas2014} but also practical ones~\cite{Ohzeki2019,Nishimura2019,Tabi2021,yarkoni2021multicar,Inoue2021,Terada2018}, a quantum annealer is regarded as a quantum solver for QUBO problems.

Pragmatically, the usability of quantum annealers for complex problems relies on their compatibility with the QUBO formulation.
Previous studies are limited to cases in which the original problem formulation has an apparent link to QUBO, such as that for combinatorial optimization problems. 
A recent study combined quantum annealing with machine learning to find the optimal arrangement of the constituent elements of a metamaterial~\cite{Kitai2020}. The original problem (optical properties of the metamaterial) was not necessarily converted to a QUBO formulation, implying the applicability of quantum annealing to general optimization problems. Specifically, they proposed a type of black-box optimization framework, in which the unknown relation between the input binary variables and the complex property values computed according to the governing equations is learned by means of a second-order regression equation and the optimal input variables are obtained using quantum annealing.

Reports of applying black-box optimization to design problems are limited to optical problems with the optimal arrangement of metamaterials described above and photonic-crystals~\cite{Inoue2022}, the structural dynamics problem of substrate vibration~\cite{Matsumori2022}, and molecular design~\cite{gao2021}.

\subsection{Design problem of noise filter}
In this study, we focus on an electric noise filter as an example electric circuit.
Noise filter performance depends on the combination of electrical components and the paths of conductors connecting them.

Products designed for electromagnetic compatibility incorporate noise filters that reduce input voltage noise to prevent high-frequency noise from affecting surrounding electronic devices. 
The electrical component allocation region needs to be determined under the constraint of a certain amount of noise attenuation, i.e., an optimal filter design is required. 
In this study, we apply black-box optimization that incorporates calculations conducted using quantum annealing to the design optimization of a noise filter that consists of two capacitors and an inductor, called a $\pi$-type filter, and demonstrate that this optimization framework is useful for electric circuit design problems. 

Topology optimization has been used for optimal design.
Although topology optimization is applicable to electric circuits~\cite{Nomura2019}, 
the inherent challenge is to avoid falling into a local optimal solution, which stems from the method being based on the gradient method.
In particular, optimization problems with many degrees of freedom related to element location, as considered in this study, generally have a complex objective function space, which can hinder the search for the global optimal solution.
The proposed optimization framework, which combines black-box optimization and quantum annealing, exploits the features of quantum annealing to avoid becoming trapped in a local optimal solution.

\subsection{Summary of contributions}
The contributions can be summarized as follows.
\begin{itemize}
\item We extend the framework of optimal design based on black-box optimization using quantum annealing to problems related to electric circuit systems.
\item We confirm that the optimization process works as an optimal design method for electric circuits by analyzing the learning process based on the relation between the number of searches and performance values.
\end{itemize}

\section{Method}
\subsection{Design problem of $\pi$-type noise filter}
\begin{figure}[t]
	\centering
	\includegraphics[width=8cm]{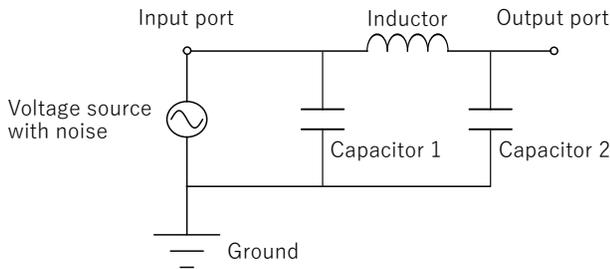}
	\caption{Circuit diagram of $\pi$-type noise filter.}
	\label{fig:filter_circuit}
\end{figure}

\begin{figure}[t]
	\centering
	\includegraphics[width=6cm]{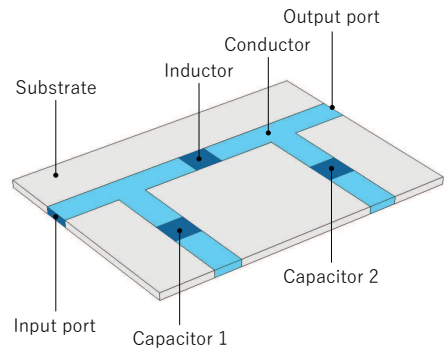}
	\caption{Example of element and conductor arrangement for $\pi$-type noise filter. The input and output ports, capacitors, and inductor are represented by simple square elements. The backplane is the electrical ground.}
	\label{fig:board}
\end{figure}

A circuit diagram of the $\pi$-type noise filter to be designed is shown in Fig. \ref{fig:filter_circuit}. 
The circuit consists of three elements, namely an inductor and two capacitors. 
Figure \ref{fig:board} shows the $\pi$-type noise filter model utilized in this study. It is assumed that the back side of the substrate is grounded. 
The performance of a noise filter is determined by the capacitance of the capacitor, the inductance of the inductor, inductive noise, and parasitic capacitance. The inductive noise and parasitic capacitance depend on the relative location of the inductor, the capacitors, and the conductor path, which does not appear in the circuit diagram but should be designed as described below.

\subsection{Black-box optimization of noise filter}
\begin{figure}[t]
\centering
\includegraphics[width=5cm]{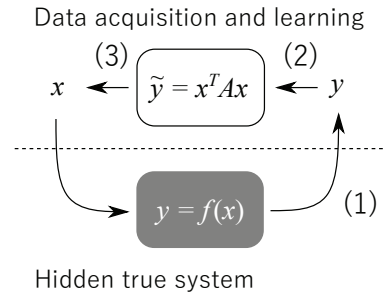}
\caption{Schematic diagram of BOCS. (1) Data $y$ for input $x$ is obtained from simulation or experiment. (2) Second-order regression equation is estimated from input $x$ and $y$. $\tilde{y}$ is estimated value. (3) Optimal $x$ is found. Here, $A$ is the coefficient of the quadratic regression equation. $f$ is an unknown function under the governing equation.}
\label{fig:bocs}
\end{figure}

The objective of black-box optimization is to obtain the input parameter $x$ that minimizes (or maximizes) the characteristic value $y$ with a small number of trials
under the condition that the relation between $x$ and $y$ ($y=f(x)$) is unknown. 
Here, we focus on Bayesian Optimization of Combinatorial Structures (BOCS)~\cite{baptista18}, which is a learning method applicable to cases where the input parameter $x$ is a binary variable, as done in the literature~\cite{Kadowaki2022}. 
In BOCS, the relation between $x$ and $y$ is learned sequentially using a quadratic regression equation of $x$.
In other words, starting with several data sets of $x$ and $y$, we (1) obtain the data $y$ for the input $x$ through simulations or experiments on a real system where the input-output relation is unknown, (2) learn the relation between data $y$ and input $x$ in quadratic form, and (3) search for the optimal input $x$ under the assumed quadratic relation. The relation between the various tasks in BOCS is summarized in Fig. \ref{fig:bocs}.

To apply this black-box optimization to the design of noise filters, we define a binary variable $x$ that specifies the location of the element and the conductor path, and employ electromagnetic field analysis using the finite element method as the data acquisition method in (1). In (3), quantum annealing is employed to find the global minimum in the regression model, which has many local minima. The solution $x$ of the quantum annealing and the corresponding output value $y$ are added to the data in the learning process. We refer to this method as BOCS-QA.
To clarify the effect of quantum annealing, a calculation using simulated annealing (BOCS-SA) instead of quantum annealing is also performed and the results are compared.

The following sections describe the binary design variables that represent electrical component positions and conductor paths and the characteristic values for evaluating filter performance.

\subsubsection{Binary design variables} \label{sec:bv}
\begin{figure}[t]
	\centering
	\includegraphics[width=9cm]{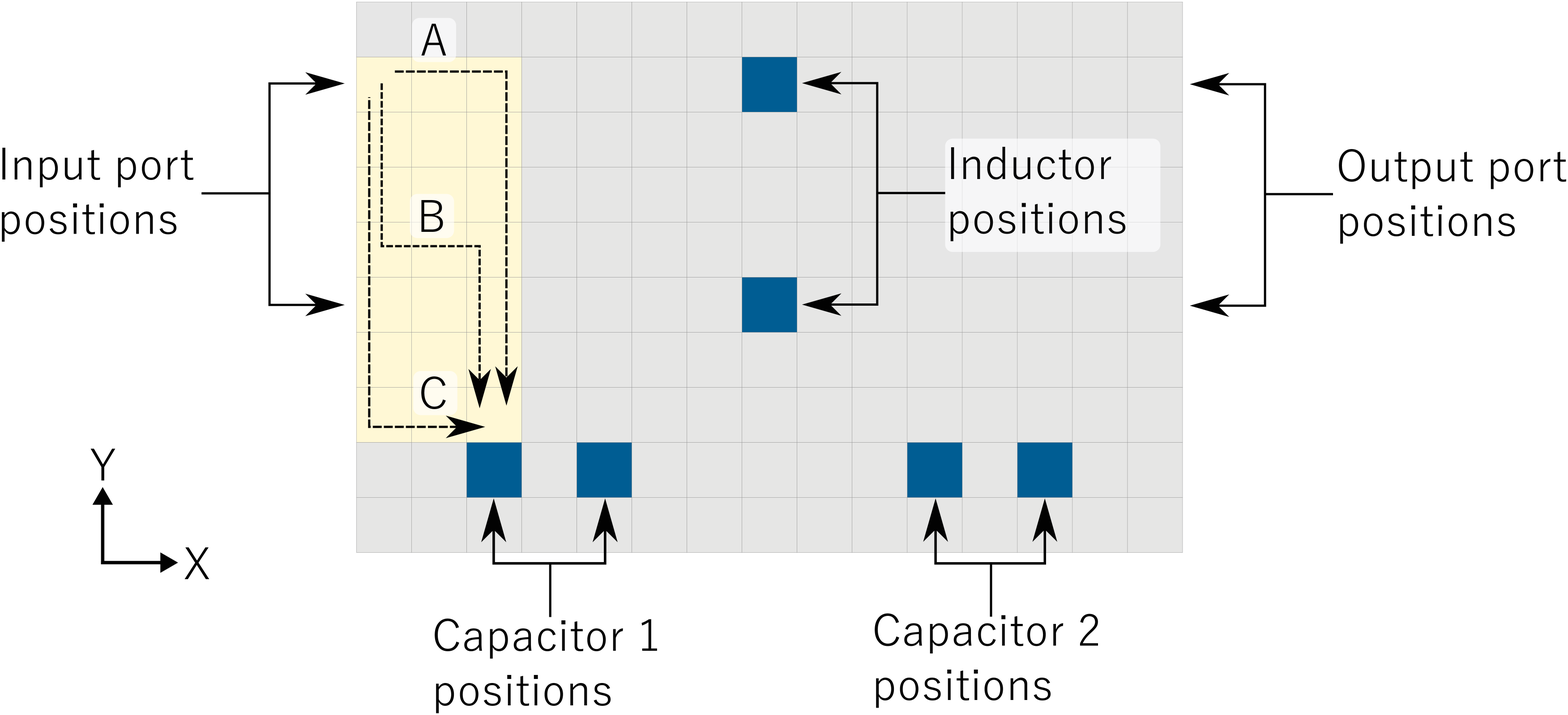}
	\caption{Candidate element positions and conductor paths. As an example of conductor paths, three candidates (A, B, and C) between the upper side of the input port and the left side of capacitor 1 are shown.}
	\label{fig:binaryx}
\end{figure}

The element positions and conductor paths between the elements are mapped to the binary variable $x$.
In this study, the problem is to select the positions of five elements (an input port, an output port, an inductor, and two capacitors) from two candidates and the conductor paths from three candidates.
In order to represent these variables as binary variables, the substrate is divided into a 10$\times$15 (X$\times$Y) grid.
The input and output ports are placed on the sides of the board and the inductor and capacitor are placed in the grid as concentrated elements, as shown in Fig. \ref{fig:binaryx}.

Three candidate paths as conductors are created by connecting the elements from top to bottom in the following manner.
\begin{enumerate}
	\item[A. ] Draw a path in the X direction and then in the Y direction.
	\item[B. ] Draw a path in the Y direction to half of the difference, then in X, and then in the remaining Y direction.
	\item[C. ] Draw a path in the Y direction and then in the X direction.
\end{enumerate}
The typical $\pi$-type noise filter, shown in Fig. \ref{fig:board}, is appropriately included as a candidate by the above conductor setting.
The present method can be simply extended to the case with more than three candidate paths if necessary.

\begin{figure}[t]
	\centering
	\includegraphics[width=8cm]{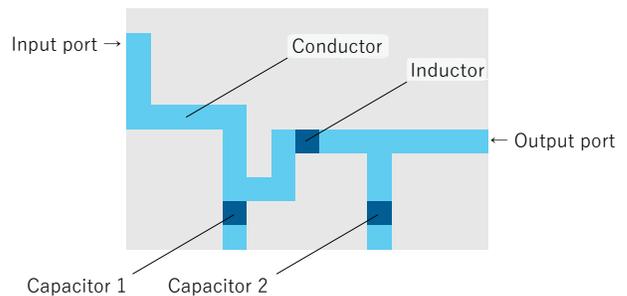}
	\caption{Circuit corresponding to bit string ``0101101010010001100100'' in one-hot representation.}
	\label{fig:example}
\end{figure}

We adopt one-hot encoding to represent noise filters in which element positions and conductor paths are selected from these candidates.
In the case considered here, 22 bits are required because there are two candidates for each of the five element positions and three candidates for each of the four conductor paths.
Let ``10'' be the state in which the element is at the bottom or on the left and ``01'' be the state in which it is at the top or on the right.
Then, let ``100'' be a conductor path that first moves in the X direction, ``010'' be one that turns in the middle, and ``001'' be one that first moves in the Y direction.
The bits that represent the conductor path follow the element position bits;
that is, the first 10 bits represent the five element positions and the latter 12 bits represent the selection of the four conductor paths. 
The bits that represent the element positions are arranged on the board in the following order from left to right: input port, capacitor 1, inductor, capacitor 2, and output port. 
The conductor paths are similarly arranged in the following order from left to right: input port - capacitor 1, capacitor 1 - inductor, inductor - capacitor 2, and capacitor 2 - output port.
For example, a circuit encoded by ``0101101010010001100100'' as binary variable $x$ is shown in Fig. \ref{fig:example}.

\subsubsection{Obtaining characteristic value}
The S-parameter $S_{21}$ is adopted as the characteristic value $y$ of the noise filter.
$S_{21}$ indicates the ratio of output power to input power.
When the input power of noise is $p_1$ and the output power is $p_2$, $S_{21}$ is expressed by the following equation,
\begin{equation}
	S_{21} = \sqrt{\frac{\lvert p_2 \rvert}{\lvert p_1 \rvert}}.
\end{equation}
We design a noise filter that minimizes $S_{21}$ under the given noise voltage. $p_1$ and $p_2$ are calculated using finite element analysis for simulating the electromagnetic field of the electric circuit model shown in Fig. 2, that is, the model in which the back of the board is the ground and the electrical components are lumped-parameter ones on the surface of the board. 
A sufficiently large air region is provided around the board in order to precisely calculate the induced noise.
A scattering boundary condition is set at the outermost boundary of the air region. 

Special procedures are required in the following two cases where the S-parameters are not correctly evaluated by the finite element method.
\begin{enumerate} 
	\item[(I)] Element position does not satisfy the one-hot constraint.
	\item[(II)] A bit in the conductor path is ``000'' (the circuit has a disconnection on the board).
\end{enumerate}
In case (I), the binary variables are unencodable to a configuration of a noise filter. Given such binary variables, instead of performing the finite element method, we calculate $y$ as a penalty according to the following formula,
\begin{equation}
	y = y_\textrm{base} + \lambda \sum_{m=1}^{5} \left( x_{2m-1} + x_{2m} -1 \right)^2, \label{eq:penalty}
\end{equation}
where $y_\textrm{base}$ is the base value of the violation of one-hot constraints, $\lambda$ is the penalty coefficient, and $x_i$ is the value of the $i$-th bit of the binary variable $x$. 
Since BOCS learns characteristic values in quadratic form, this penalty of one-hot constraints is also expected to be learned.

\begin{figure}[t]
	\centering
	\includegraphics[width=5cm]{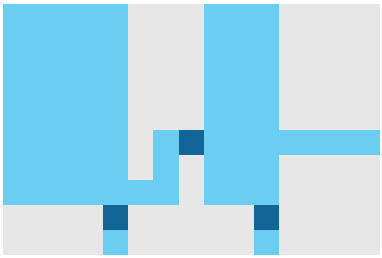}
	\caption{Circuit corresponding to bit string ``1001011001000001000100''. The conductor paths between the input power port and capacitor 1 and those between the inductor and capacitor 2 are not selected. To avoid disconnection, conductors spread over the board are assigned.}
	\label{fig:disconne}
\end{figure}

In case (II), a meaningful S-parameters  for evaluating a noise filter performance cannot be obtained because the conductor path is disconnected such that voltage is conducted neither from a normal signal nor noise.
We assign a dummy conductor that avoids the disconnection, as shown in Fig. \ref{fig:disconne}.
Note that when multiple conductor paths are selected, such as ``011'', we take the sum of the conductor paths.

To summarize, we calculate the characteristic value of a noise filter $y$ using the following equation,
\begin{align}
    z &\equiv \sum_{m=1}^{5} \left( x_{2m-1} + x_{2m} -1 \right)^2, \\
    y &= 
    \begin{cases}
    & S_{21} \hskip 0.5cm \textrm{( for } z = 0 \textrm{ )},\\
    & y_\textrm{base} + \lambda z \hskip 0.5cm \textrm{( for } z \neq 0 \textrm{ )}.
    \end{cases} 
\end{align}

\subsection{Parameters for circuit model and black-box optimization}
For the calculation of characteristic values, the substrate thickness, width, and height are set to 1.6, 150, and 100 mm, respectively. An air area of 30 mm is provided around the board. Scattering boundary conditions are set at the outermost boundaries of this air region.
The substrate is divided into a 10$\times$15 grid, as introduced in 
 section \ref{sec:bv}.

The physical constants of the power supply port, capacitor, and inductor are set to 50 \si{\ohm}, 100 \si{\farad}, and 10 \si{\henry}, respectively. 
The substrate's relative permittivity, relative permeability, and conductivity are set to 4.5, 1, and 1.0 $\times$ 10$^{-8}$ \si{\siemens}/m, respectively, assuming an FR-4 substrate.
The conductor is treated as a perfect conductor.
In addition, $S_{21}$ was calculated using a frequency analysis at 10 M\si{\hertz}.
For Eq. (\ref{eq:penalty}), we set $y_\textrm{base} = -60$ and $\lambda=10$.

The quantum annealer was Advantage\_system4.1 by D-Wave Systems.
We adopted the Python library dwave-neal by D-Wave Systems as a simulated annealing method. 
The sampling number was set to 3000 when solving the problem. The $x$ value that gave the smallest $y$ was adopted as the next candidate.
For the initial training datasets of BOCS, we prepared 20 randomly generated binary variables $x$ and their corresponding characteristic values $y$. 
BOCS-QA and -SA were performed until 300 searches were conducted.

\section{Results and Discussion}
We compare the results of the BOCS-QA and BOCS-SA calculations with those of random search in which binary variables were randomly generated.

\begin{figure}[t]
	\centering
	\includegraphics[width=7cm]{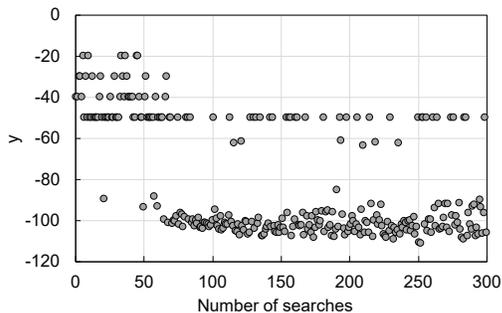}
	\caption{Full search history of BOCS-QA.}
	\label{fig:bocsqa-all}
\end{figure}

\begin{figure}[t]
	\centering
	\includegraphics[width=7cm]{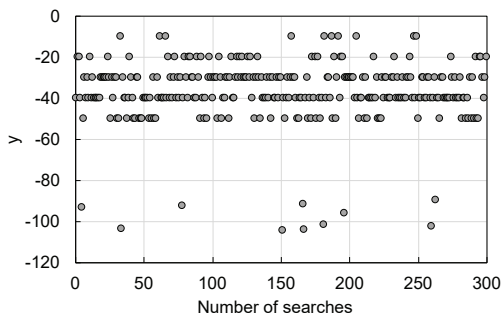}
	\caption{Full search history of random search.}
	\label{fig:random-all}
\end{figure}

First, the results of all search histories of BOCS-QA and random search are shown in Figs. \ref{fig:bocsqa-all} and \ref{fig:random-all}, respectively.

The learning processes of BOCS-QA and random search are different. Figure~\ref{fig:bocsqa-all} shows that BOCS-QA mainly learned the penalty term in Eq. (\ref{eq:penalty}) in the beginning (before approximately 60th search), and subsequently started to learn on the bases of the performance of the noise filter $S_{21}$, suggesting that the design of the penalty term facilitated learning.
Then, the highest record of $S_{21}$ was steadily set.
On the other hand, the random search shown in Fig. \ref{fig:random-all} searched for a feasible noise filter in very rare cases. There is no particular trend.
The number of solutions that satisfy the one-hot constraint is ten, which is close to the expected value. The probability that a random binary variable satisfies the one-hot constraint is $2^5 / 2^{10} = 1 / 32$, so the expected number for 300 searches is nine.


\begin{figure}[t]
	\centering
	\includegraphics[width=8cm]{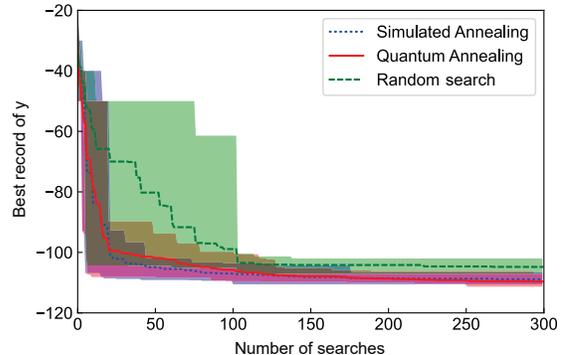}
	\caption{Updated records of $y$. The solid and dotted lines represent the mean and the filled area represents the area between the maximum and minimum values.}
	\label{fig:variance_results}
\end{figure}

The update records of the characteristic value $y$ versus the number of searches are shown in Fig. \ref{fig:variance_results}.
Since BOCS-QA, BOCS-SA, and random search are randomized algorithms, the mean, minimum, and maximum values were calculated for ten trials.
BOCS-QA and BOCS-SA steadily search for a noise filter with good performance, whereas random search tends to have a large variance (especially with a small number of searches).
The steady performance improvement of BOCS-QA and BOCS-SA shown in Fig. \ref{fig:variance_results} is due to the successful learning of $S_{21}$, as confirmed in Fig. \ref{fig:bocsqa-all}.
At 300 searches, BOCS-QA shows slightly better performance than that of BOCS-SA in terms of the average, minimum, and maximum values, as shown in Table \ref{table:results}.

\begin{figure}[t]
	\centering
	\includegraphics[width=8cm]{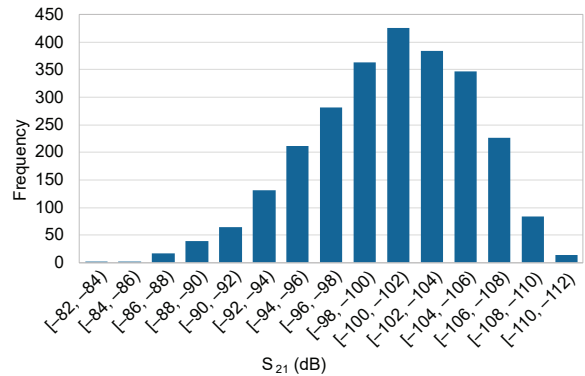}
	\caption{Histogram of $S_{21}$ value in decibels when element positions are specified uniquely and there is one conductor between elements.}
	\label{fig:filter_histgram}
\end{figure}

Next, we evaluate the filter performance of the obtained solution.
Since there are $2^{22}$ cases (expressed in 22 bits), enumerating the performance of all solutions is unrealistic. We therefore choose only the relevant solutions with unique element positions and a single conductor path between elements. This gives a total of 2592 cases ($2^5 = 32$ combinations of element positions and $3^4 = 81$ combinations of conductor positions).
Figure \ref{fig:filter_histgram} shows a histogram of the $S_{21}$ value in decibels.
For our settings, noise filters whose $S_{21}$ is under $-108$dB are rare (approximately 3\%).
Since the average records of BOCS-QA and BOCS-SA are in the top 0.8\% and 1.9\%, respectively, as shown in Table \ref{table:results}, these methods finding such filters in 300 searches are considered efficient.
\begin{table}
	\caption{Comparison of results obtained by various methods.}
	\centering
	\begin{tabular}{c|ccc}\hline
	Method & Object & Value & Rank \\ \hline
	\multirow{3}{*}{QA} & Best & $-111.34$ dB & 1st \\
	& Average & $-109.64$ dB & 19th \\
	& Worst & $-106.97$ dB & 192nd \\ \hline
	\multirow{3}{*}{SA} & Best & $-110.55$ dB & 14th \\
	& Average & $-108.91$ dB & 48th \\
	& Worst & $-104.80$ dB & 528th \\ \hline
	\multirow{3}{*}{Random} & Best & $-107.12$ dB & 180th \\
	& Average & $-104.80$ dB & 192th \\
	& Worst & $-102.00$ dB & 1058th \\ \hline
	\end{tabular} \label{table:results}
\end{table}

\begin{figure}[t]
	\centering
	\includegraphics[width=5cm]{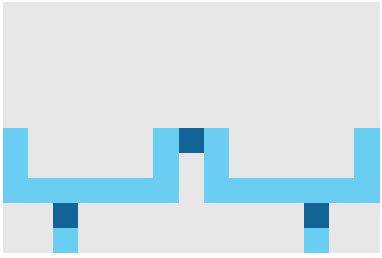}
	\caption{Noise filter obtained by BOCS-QA.}
	\label{fig:best_filter}
\end{figure}

The configuration of the best-performing noise filter obtained using BOCS-QA is shown in Fig.~\ref{fig:best_filter}.
In this case, the value of $S_{21}$ was $-111.34$ dB.
The input port and capacitor are placed close to each other, preventing performance degradation due to induced noise. This shows that the obtained configuration is physically reasonable.

In this study, we formulated a problem with two candidates for the element positions and three candidates for the conductor paths. If we considered a large-scale problem with a larger number of candidates, the probability of finding a well-posed noise filter by chance using random search would be much smaller and the superiority of BOCS-QA and BOCS-SA would be more significant.

\section{Conclusion and Outlook}
To find input parameters that provide the desired characteristics with a small number of searches, we proposed an iterative optimization method that incorporates quantum annealing in the BOCS framework and applied it to the problem of designing noise filters.
A $\pi$-type noise filter that consists of two capacitors and an inductor was considered.
A model was created to select two candidates for the location of these elements and three candidates for the path of the conductor connecting the elements.
The results show that a high-performance noise filter can be efficiently found and that the search progresses more stably than does random search. 
This shows that the framework that incorporates quantum annealing into black-box optimization is applicable to electric circuit design problems.
The present method could help engineers meet the high demand for electrical products. 

Beyond the optimization of electric components demonstrated here, system-level optimization of electric devices is a topic for future work. 
It could lead to multiphysics optimal design that requires simultaneous optimizations of multiple phenomena.

The proposed BOCS framework was proven to work with quantum annealing and simulated annealing. A comparison of these two versions showed only a slight difference.
A recent study that compared the two solvers in an black-box optimization framework
also concluded that clear performance improvements using quantum annealing are rare~\cite{Matsumori2022}. However, a clear advantage of quantum annealing in finding optimal solutions, achieved by adjusting the annealing schedule, has recently been reported~\cite{Koh2022}.
Future research should thus examine in detail the scheduling protocols to further improve the performance of BOCS with quantum annealing. In addition, a recent improvement of the learning process~\cite{Kadowaki2022} could be integrated into the present BOCS framework to speed up the whole optimization process.

\bibliographystyle{apsrev4-1}
\bibliography{NoiseFilter_1226}

\end{document}